\documentstyle[editedvolume,epsf]{crckapb}

\begin{opening}
\title{Aspects of the dynamics underlying solar and stellar dynamo models}

\author{Reza Tavakol, Eurico Covas}
\institute{Astronomy Unit\\
School of Mathematical Sciences \\
Queen Mary \& Westfield College \\
Mile End Road \\
London E1 4NS, UK
}
\author{David Moss}
\institute{Department of Mathematics \\
The University \\
Manchester M13 9PL, UK}
\end{opening}

\runningtitle{Aspects of dynamics}

\begin{document}
\begin{abstract}
Observations of the Sun and solar-type stars continue to reveal
phenomena whose understanding is very likely to require a nonlinear
framework.  Here we shall concentrate on two such phenomena, namely the
grand minima type behaviour observed in the Sun and solar-type stars
and the recent dynamical variations of the differential rotation in the
solar convection zone, deduced from the helioseismic observations, and
discuss how their explanations have recently motivated the
development/employment of novel ideas from nonlinear dynamics.
\end{abstract}
\section{Introduction}
An important characteristic of the Sun and solar-type stars
is their variability over a very large range of time (and space) scales,
including the intermediate time scales of
$\sim 10^0 - 10^4$ years (Weiss 1990).
Though ultimately spatiotemporal in character,
these variabilities fall into two categories: (i) those
whose main features can be explained in terms of temporal models
and (ii) those which require a spatiotemporal approach for their explanations.

Here we shall be
focusing on two such forms of variability, namely
the grand minima type episodes seen in in sunspot activity and
the recently observed dynamical variations of the differential rotation
in the solar convection zone.

Evidence for the presence of the grand--minima type behaviour
comes from
a variety of sources, including
the studies of the
historical records of the annual mean sunspot data since 1607 AD which show the
occurrence of epochs of suppressed sunspot activity, such as the {\it Maunder
minimum} (Eddy 1976, Foukal 1990, Wilson 1994, Ribes \& Nesme-Ribes 1993,
Hoyt \& Schatten 1996). Further research, employing $^{14}C$
(Stuiver \& Quey 1980; Stuiver \& Braziunas 1988, 1989) and $^{10} B$ (Beer
{\em et al.} 1990, 1994a,b, Weiss \& Tobias 1997) as proxy indicators, has
provided additional strong evidence that the occurrence of such epochs of reduced
activity (referred to as {\it grand minima}) has persisted with
irregular time intervals in the past.
Further, they appear to be typical of solar-type stars (Baliunas {\em
et al.} 1995).

Evidence for the variations in the dynamical behaviour of the differential rotation
comes from
recent analyses of the inversions of the helioseismic
data from both the
Michelson
Doppler Imager (MDI) instrument on board the SOHO
spacecraft and the Global Oscillation Network Group (GONG)
project
which provide evidence to show
(i) that
the previously observed
surface torsional oscillations with periods
of about 11 years (e.g.\ Howard \& LaBonte 1980; Snodgrass, Howard \& Webster 1985;
Kosovichev \& Schou 1997; Schou {\em et al.} 1998)
extend significantly  downwards
into the solar convective zone to depths of
about 10 percent in radius
(Howe {\em et al.} 2000a and Antia \& Basu 2000;
see Covas {\em et al.} 2000a for a recent study),
and
(ii) that
the dynamical regimes at the base
of the convection zone may be different from those
observed at the top, having either significantly shorter
periods (Howe {\em et al.} 2000b) or non--periodic behaviour
(Antia \& Basu 2000).

Our aim here is to briefly describe some recent attempts
at understanding these phenomena within a nonlinear theoretical
framework.

\section{Intermittency as a mechanism underlying grand minima type behaviour}
The seemingly irregular variations of the grand minima type in the
sunspot activity are of
interest for at least two reasons.
Firstly, they are theoretically challenging, especially in
view of the absence of any naturally occurring mechanisms in
the Sun and solar-type stars
with appropriate time scales (Gough 1990).
Secondly, given their time scales,
such variations can be of potential consequence for the occurrence of climatic
variability on similar time scales (e.g.\ Friis-Christensen \& Lassen 1991,
Beer {\em at al.} 1994b, Lean 1994, Stuiver, Grootes \& Braziunas 1995,
Baliunas \& Soon 1995, Butler \& Johnston 1996,
White {\em et al.} 1997).

These considerations have motivated
considerable
effort into trying to understand the
mechanism(s) underlying such variations, employing a variety of approaches
(e.g.\ Weiss {\em et al.} 1984; Sokoloff and Nesme--Ribes 1994; Tobias
{\em et al.} 1995; Knobloch and Landsberg 1996; Schmitt {\em et al.} 1996, Knobloch {\em et al.} 1998;
Tobias 1998a,b; Tobias {\em et al.} 1999).

In principle, there are essentially two frameworks within which such
variabilities could be studied: stochastic and deterministic.
In practice, however, it is difficult to distinguish between these two frameworks (Weiss
1990). Here we shall concentrate on the deterministic approach and recall
that whatever the relevant framework may turn out to be,
the deterministic components will still be
present and are likely to play an important role.

Within this framework, a number of approaches have
been adopted. Among these is
the employment
of low dimensional ODE models that are obtained using the Normal Form
approach (Spiegel 1994, Tobias {\em et al.} 1995, Knobloch {\em et al.} 1996).
These have led to robust modes of behaviour which are
of potential importance in accounting
for certain aspects of  solar variability
of the grand minima type,
such as various forms of
amplitude modulation of the magnetic field
energy.

An alternative way to understand this
type of variability in the sunspot record
is to postulate that this type
of behaviour is caused by some
form of dynamical
intermittency, an idea that in various forms goes back at least to the late 1970s (e.g.
Tavakol 1978, Ruzmaikin 1981, Zeldovich {\em et al.} 1983, Weiss {\em et al.}
1984, Spiegel 1985, Feudel {\em et al.} 1993, Schmitt {\em et al.} 1996, Brooke 1997, Tworkowski {\em et al.} 1998).

Much effort has gone into producing evidence for the
occurrence of various forms of dynamical intermittency
in both truncated and PDE mean-field dynamo models.
Given the large number of intermittency types that
dynamical systems theory has produced and bearing in mind
the complexity of the full dynamo equations,
a natural approach would be
to single out the main generic ingredients of stellar dynamos
and to study their dynamical consequences.
For axisymmetric dynamo
models, these ingredients consist of the presence of invariant subspaces,
non-normal parameters and the non-skew property (see e.g. Ashwin {\em et al.} 1999 for
definitions).
The dynamics underlying such
systems has recently been studied (Covas {\em et al.} 1997c, 1999b; Ashwin
{\em et al.} 1999), leading to a number of novel
phenomena, including a new type of intermittency, referred to as
{\it in--out intermittency} (see also Brooke 1997).

The easiest way to characterise in--out
intermittency is by contrasting it with on--off intermittency. Briefly, let
$M_I$ be the invariant subspace and $A$ the attractor which exhibits either
on--off or in--out intermittency. One then has on--off or in--out
intermittencies, depending respectively upon whether
the intersection $A_0=A\cap M_I$ is a
minimal attractor or not. The name in--out is chosen
to be indicative of the fact that
in this case
there can be different invariant sets in $A_0$ associated with the attraction and
repulsion transverse to $A_0$.
Another crucial
difference between the two is that, as opposed to on--off intermittency, in the
case of in--out intermittency the minimal attractors in the invariant
subspaces do not necessarily need to be chaotic and hence the trajectories
can (and often do) shadow a periodic orbit in the `out' phases (see Ashwin {\em et al.}
1999 for details).

\begin{figure}[!htb]
\centerline{\def\epsfsize#1#2{0.7#1}\epsffile{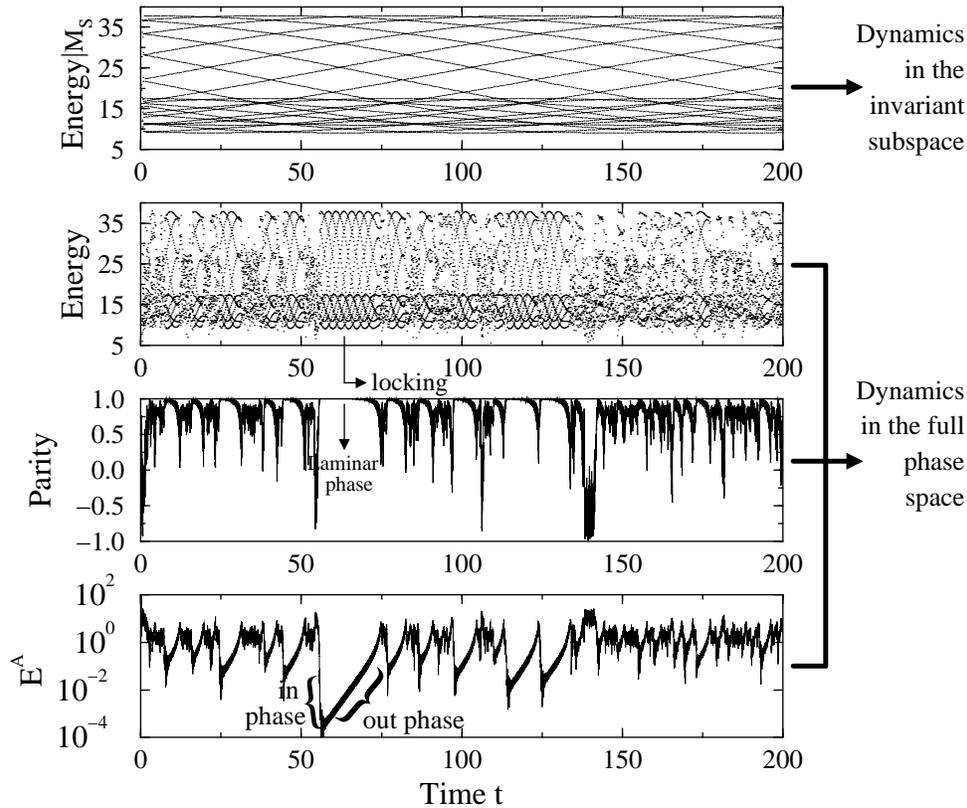}}
\caption{ \label{inoutpde2}
In--out intermittency in the axisymmetric PDE mean--field dynamo model.
The parameters used were $r_0=0.4$, $C_{\alpha}=1.505$, $C_{\omega}=-10^{5}$,
$f=0.7$, together with an algebraic form of $\alpha$.
To enhance visually the periodic locking we have time sampled the
series in the two upper panels.
See Covas {\em et al.} (2000d) for details of the model.
}
\end{figure}

Given the genericity of the ingredients required for the presence of this type
of intermittency, it is of interest to see if this type
of behaviour can indeed occur in the context of dynamo models.
Interestingly, this form of intermittency has been concretely shown to
occur in a number of dynamo models, including both
ODE (Covas et al. 1997c) and PDE (Covas et al. 2000c) mean field dynamo models.
An example of such an occurrence is given in
Fig.\ \ref{inoutpde2}, where in--out intermittency occurs
for an axisymmetric PDE mean--field dynamo model
(See Covas {\em et al.} (2000d) for details).

We should add here that in addition to in--out intermittency,
firm evidence has also been
found for the occurrence of other forms of intermittency
in mean field dynamo models. These include
crisis (or attractor merging) intermittency, in both
ODE
(Covas \& Tavakol 1997) and PDE (Covas \& Tavakol 1999)
dynamo models, and
Type I Intermittency, in both ODE (Covas {\em et al.} 1997c)
and PDE models (Covas \& Tavakol 1999). We further note that
on-off intermittency (Platt {\em et al.} 1993, see also
Schmitt {\em et al.} 1996 and Ossendrijver 2000) is also likely
to be relevant for such models, given their invariant
subspaces.

The demonstration of the occurrence of
various form of intermittency establishes an
important element of the {\it multiple intermittency hypothesis}
(Tavakol \& Covas 1999), according to which the grand minima type variability
in solar-type stars may be
understood in terms of one or a number of types of
dynamical intermittency.
What remains
to be shown is whether these types of intermittency
still persist in more realistic models and ultimately,
through detailed comparison
with observational and proxy
data, whether they actually occur in
the Sun and stars,
and whether they can in fact
account for the grand minima type behaviour.

The important point to be emphasised here
is that in addition to theoretical predictions
for these types of intermittency, their occurrence has been
conclusively demonstrated not only in truncated dynamo models,
but also in non-truncated PDE dynamo models (in three spatial dimensions,
although so far restricted to axisymmetry, and time).

\section{Spatiotemporal bifurcations as a
mechanism
for different modes of behaviour in
the solar convection zone}
An important outcome of the recent inversion of the helioseismic data
has been to provide increasing evidence to show that
the Sun is likely to be very active dynamically throughout
the convection zone.
To establish firmly the precise nature of
these variations, future observations are
clearly required. However, despite
the error bars that such inversions are bound
to entail, the results so far seem to point
to the very interesting possibility that
the variations in the differential rotation
can have different periodicities/behaviours
at different depths in the solar
convection zone, having either markedly reduced periods (Howe {\em et al.} 2000b)
or non--periodic behaviour (Antia \& Basu 2000) at the bottom of the convective zone.

There are in principle two ways to account for such behaviours.
One could imagine different physical mechanisms at different
physical locations in the convection zone, resulting in
different dynamical behaviours in those locations. Alternatively,
a dynamical mechanism could be sought that
could produce such different behaviours at different spatial locations,
without requiring
different physical mechanisms at those locations.

As an example of the latter mechanism, spatiotemporal
fragmentation/bifurcation has recently been
proposed as a possible natural dynamical mechanism to
account for such observed multi-mode
behaviours in different parts of the
solar convection zone (Covas {\em et al.} 2000b).
In this way, dynamical regimes with different temporal behaviours
can coexist at different spatial locations,
for certain {\it given} values of the control parameters of the system.
The important point here is that,
in contrast to the usual
temporal bifurcations, which result in
identical temporal behaviour at each spatial
point, and which require changes in parameters in order to be initiated,
spatiotemporal bifurcations can result in different
dynamical modes of behaviour at different locations
without requiring changes in the
control parameters.
Also, importantly, the occurrence of
such diverse modes of behaviour does not require
different physical mechanisms at different locations.

Evidence for the occurrence of
this mechanism was found in the context of
a two dimensional axisymmetric mean field
dynamo model operating in a spherical shell, with
a semi--open outer boundary condition,
in which the only nonlinearity  is the action
of the azimuthal component of the Lorentz force of the
dynamo generated magnetic field on the solar
angular velocity (Covas {\em et al.} 2000b).
The
zero order angular velocity was
chosen to be consistent with the most recent helioseismological data (MDI).
Subsequently it has been shown, through a detailed study (Covas {\em et al.} 2000c),
that the occurrence of this type of behaviour
does not depend upon the details of the model employed,
thus providing strong evidence to support the idea that spatiotemporal fragmentation
is likely to occur in general dynamo settings.

\begin{figure}[!htb]
\centerline{\def\epsfsize#1#2{0.6#1}\epsffile{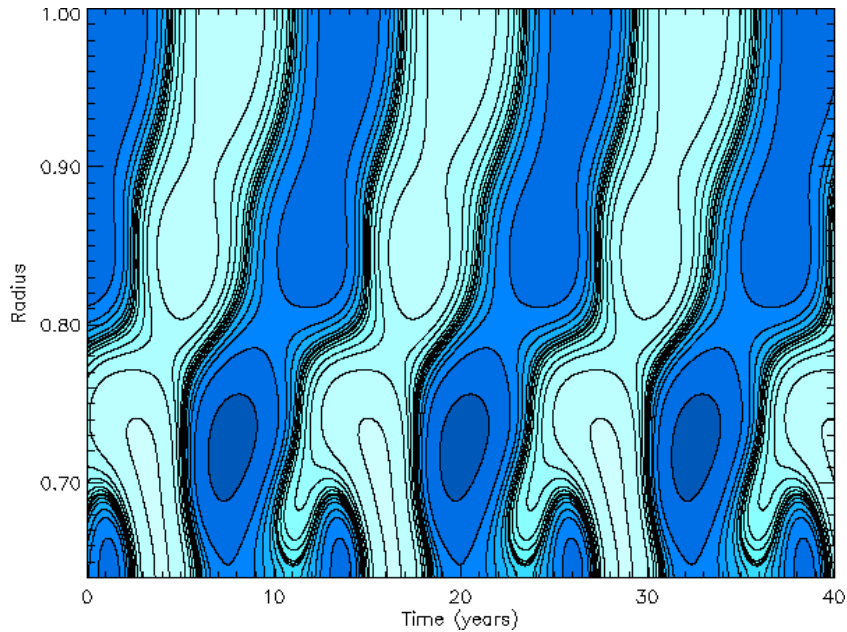}}
\caption{\label{velocity_radial_latitude=30.Pr=0.90}
Radial contours of the angular velocity residuals $\delta \Omega$
as a function of time for a cut at latitude $30^{\circ}$.
Parameter values
are  $R_{\alpha}=-11.0$, $P_r=0.9$,
$R_{\omega}=44000$ for $\alpha(r,\theta)=\alpha_r\cos\theta\sin^2\theta$.
See Covas {\em et al.} (2000b) for the details of the model.
}
\end{figure}

As an example of this type of spatiotemporal fragmentation/bifurcation,
we have plotted in Fig.\ \ref{velocity_radial_latitude=30.Pr=0.90}
the radial contours of the
angular velocity residuals $\delta \Omega$,
as a function of time for a cut at a fixed latitude.
This demonstrates how, as a result of such spatiotemporal bifurcation,
the period (as well ads the phase) of the angular velocity residuals
vary going from the top to the bottom of the convection zone.
The details of the model are given in Covas {\em et al.} (2000b,c).

\begin{figure}[!htb]
\centerline{\def\epsfsize#1#2{0.6#1}\epsffile{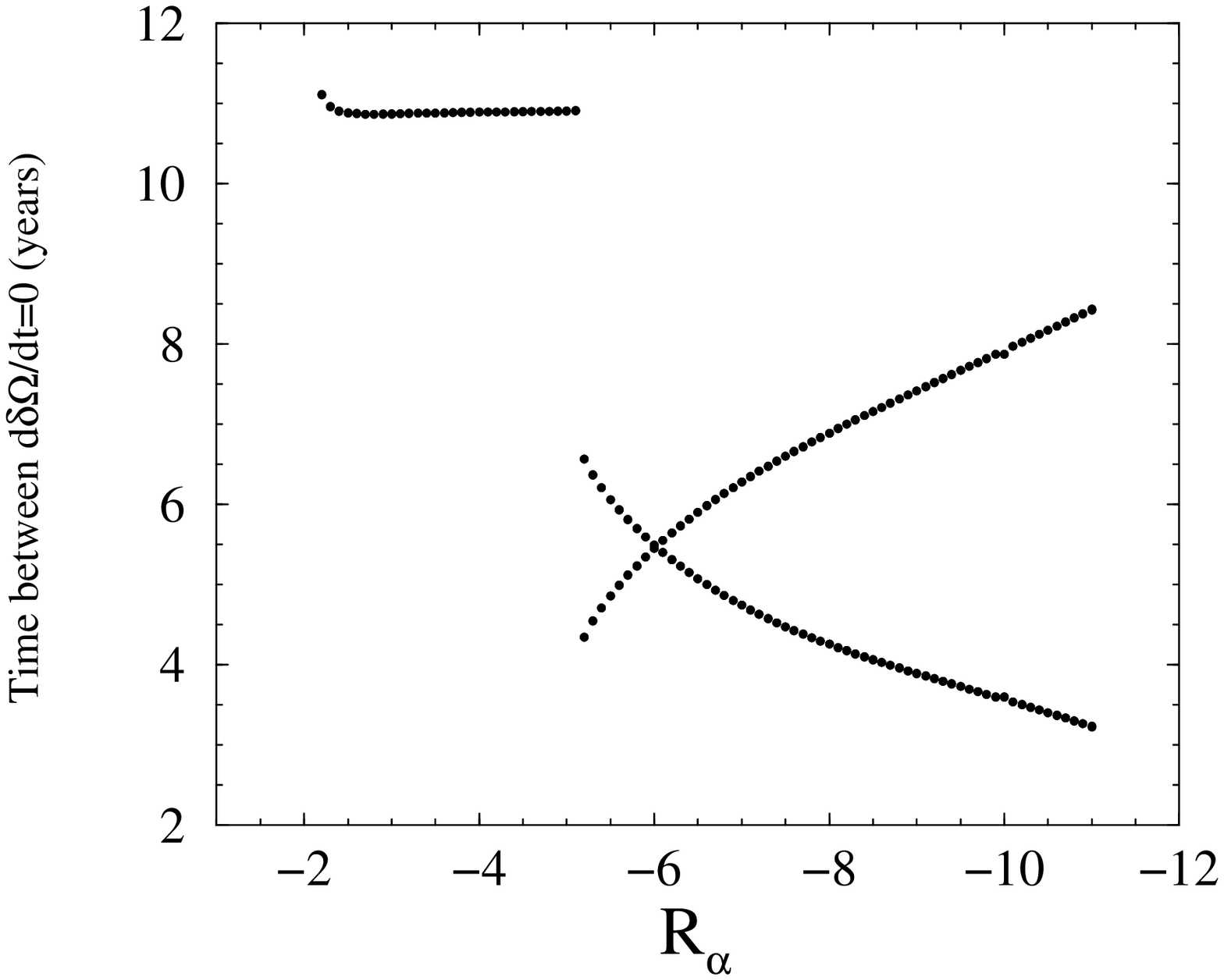}}
\caption{\label{periods.bifurcations.Calpha}
Bifurcation diagram showing the time between minima of the curve
$\Omega(r,\theta,time)$. It shows clearly a sudden bifurcation from
a fundamental 11 year cycle to a cycle with two timescales (with sum
11 years).
Parameter values
are  $R_{\alpha}=-11.0$, $P_r=1.0$,
$R_{\omega}=44000$ for a $\alpha(r,\theta)=\alpha_r\cos\theta\sin^2\theta$.
See Covas {\em et al.} (2000b) for the details of the model.
}
\end{figure}

To show the the presence of the spatiotemporal bifurcation
more clearly, we also studied the
bifurcation diagram by calculating the time between the minima of
$\delta \Omega(r=0.66 R_{\odot},\theta=30^{\circ})$,
the perturbation to the zero order rotation rate at a fixed latitude and radius,
as a function of time.

Fig.\ \ref{periods.bifurcations.Calpha} shows an example of such bifurcation,
demonstrating clearly a sudden bifurcation from
a fundamental 11 year cycle to a cycle with two periods
(whose sum is 11 years).
Thus this mechanism is capable of producing different modes of
behaviour at different locations, for given values of the control parameters
of the system.

\section{Discussion}

We have briefly discussed a number of recent attempts
at understanding two dynamical features observed in the Sun (and
solar type stars), in terms of a number of novel concepts from
nonlinear dynamics. A crucial feature of these
types of explanation is that they are based on
the nonlinear nature of the regimes under study,
rather than on any specific physical mechanisms
possessing the required time scale
variabilities. This is important, particularly in view of the
fact that much remains unknown about the
underlying physics in these regimes.
\\

We would like to thank Peter Ashwin, Axel Brandenburg, John Brooke,
and Andrew Tworkowski for the work we have done together.
EC is supported by a PPARC fellowship.

\end{document}